\documentclass[10pt]{amsart}
\usepackage{amsmath, amsthm, amssymb}
\usepackage{graphicx}
\usepackage{mathbbol}
\usepackage{txfonts}
\usepackage[usenames]{color}

\newtheorem{thm}{Theorem}[section]
\newcommand{\bt}{\begin{thm}}
\newcommand{\et}{\end{thm}}

\newtheorem{cor}[thm]{Corollary}   

\newcommand{\bc}{\begin{cor}}
\newcommand{\ec}{\end{cor}}

\newtheorem{lem}[thm]{Lemma}   

\newcommand{\bl}{\begin{lem}}
\newcommand{\el}{\end{lem}}

\newtheorem{prop}[thm]{Proposition}
\newcommand{\bp}{\begin{prop}}
\newcommand{\ep}{\end{prop}}

\newtheorem{defn}[thm]{Definition}

\newcommand{\bd}{\begin{defn}}    
\newcommand{\ed}{\end{defn}}

\newtheorem{rmrk}[thm]{Remark}   

\newcommand{\br}{\begin{rmrk}}
\newcommand{\er}{\end{rmrk}}

\newcommand{\be}{\begin{equation}}
\newcommand{\ee}{\end{equation}}

\newcommand{\mass}{{\mathbf M}}

\def\iff{\Longleftrightarrow}
\def\implies{\Longrightarrow}

\begin{document}

\title[$\mathcal{SF}$-Convergence]
{Spacetime Intrinsic Flat Convergence}

\author{Christina Sormani}
\thanks{NSF DMS - 1309360 and Oberwolfach Workshop ID 1832: Mathematical General Relativity}
\address{}
\email{}

\date{}


\begin{center} \Large{Oberwolfach Report\\Spacetime Intrinsic Flat Convergence}
\\
.
\\
Christina Sormani \footnote{NSF DMS - 1309360 and Oberwolfach Workshop ID 1832: Mathematical General Relativity}
\\
.
\\
\end{center}

The intrinsic flat ($\mathcal{F}$) convergence of Riemannian manifolds was introduced by the author jointly with Stefan Wenger in \cite{SW-JDG}.   The notion was first applied to General Relativity jointly with Dan Lee in \cite{Lee-Sormani}.  Since then it has been applied in work of Lan-Hsuan Huang, Jeff Jauregui, Dan Lee, Philippe LeFloch, Anna Sakovich, and Iva Stavrov to prove the $\mathcal{F}$-stability of
special cases of the Positive Mass Theorem, the Hyperbolic Positive Mass Theorem and the Penrose Inequality 
\cite{Lee-Sormani-2}\cite{HLS}\cite{Jauregui-lower}\cite{LeFloch-Sormani}\cite{Sakovich-Sormani-H}\cite{Sormani-Stavrov}.
.    

Shing-Tung Yau has suggested that one develop a similarly useful notion of convergence for sequences of spacetimes.  One might then apply it to answer the following questions: 
\begin{itemize}
\item {\em What does it mean to say the universe is approximately an Friedmann-Lema\^itre-Robertson-Walker big bang spacetime when it has gravity wells and black holes? } 
\item {\em In what sense is one maximal development close to another if they have approximately the same initial data but the control on initial data is not strong enough to prevent gravitational collapse? } 
\item {\em In what sense is a black hole spacetime of small mass close to Minkowski space?}
\end{itemize}

First recall that stronger notions of convergence are not suited to questions where long thin gravity wells can develop.  The development of such wells in sequences of Riemannian manifolds prevent smooth, Lipschitz, uniform, and Gromov-Hausdorff (GH) convergence.  However $\mathcal{F}$-convergence was designed specifically so that Ilmanen's examples of a sequence of spheres with wells converges.   Under $\mathcal{F}$ convergence, the wells disappear in the limit.  Indeed all regions of small volume disappear and sequences of manifolds whose volume converge to $0$ disappear as well.  If the sequence does not disappear, the intrinsic flat limit is an {\em integral current space}, $(X,d,T)$.  In particular the limit space is a metric space endowed with a biLipschitz collection of charts.  

Recall that Federer-Flemming developed the notion of integral currents to extend the notion of submanifolds and solve Plateau's problem.  Integral currents, $T$, act on $m$ forms, $omega$, via integration and have boundaries, $\partial T(\omega)=T(d\omega)$, and integer weighted volumes, $\mass(T)$. 
Federer-Flemming define the flat distance between pairs of integral currents:
\be
d_F(T_1, T_2) = \inf\{ \mass(A) + \mass(B):\, A+\partial B = T_1-T_2 \}.
\ee
So this is intuitively measuring the volume between the generalized submanifolds.   Ambrosio-Kirchheim extended this entire theory to metric spaces in \cite{AK}.  

Wenger and I then defined the $\mathcal{F}$ distance:   
\be\label{old-d}
d_{\mathcal{F}}((X_1, d_1, T_1), (X_2, d_2, T_2)) = \inf\left\{d_F^Z(\varphi_{1\#}T_1, \varphi_{2\#}T_2)
: \,\, \varphi_i: X_i \to Z \right\}
\ee
where the infimum is taken over all distance preserving maps, $\varphi_i: X_i \to Z$, and over all complete metric spaces, $Z$, of the flat distance, $d_F^Z$, between the pushforwards of the integral current structures, $\varphi_{i\#}T_i$ \cite{SW-JDG}.   In the same paper we proved Ilmanen's sequence of spheres with increasingly many increasingly thin wells converges to the sphere by constructing an explicit metric space $Z$ in one dimension higher which filled in all the wells.    In work with Sajjad Lakzian we proved $\mathcal{F}$ convergence for sequences of manifolds converging smoothly away from singular sets when there are volume, area, and distance bounds around those singular sets \cite{Lakzian-Sormani}.  

When trying to extend $\mathcal{F}$ convergence to a spacetime intrinsic flat convergence, we first encountered the difficulty that Lorentzian manifolds do not have a metric space structure and so in (\ref{old-d}) we had no notion of distance preserving maps, $\varphi_i$ : 
$d_Z(\varphi_i(p), \varphi_i(q))=d_{X_i}(p,q)$.  If $\varphi_i$ were replaced by Riemannian isometries then the $d_{\mathcal{F}}$ would always be $0$.  But how would one possibly define anything but a Lorentzian isometry between spacetimes where there is no external distance structure to speak of?  Lars Andersson suggested that we use a canonical time function like the cosmological time function, $\tau$, to create a Riemannian manifold by adding twice $d\tau^2$ to the Lorentzian metric.  Then define the spacetime intrinsic flat distance as the $\mathcal{F}$ distance between the two Riemannian manifolds and somehow keep track of the causal structure.  After a few years, trying to deal with the singularities that could arise where $\tau$ was not smooth, this approach was abandoned.

Carlos Vega and I then decided to convert a spacetime directly into an integral current space rather than a Riemannian manifold \cite{Sormani-Vega-null}.  Taking any time function, $\tau$, on our spacetime, we defined the null distance as follows:
\be
\hat{d}_\tau(p,q)=\inf_\beta \hat{L}_\tau(\beta)=\inf_\beta \sum_{i=1}^k |\tau(\beta(t_i))-\tau(\beta(t_{i+1})|
\ee
where the inf is over all piecewise causal curves $\beta$ from $p$ to $q$, which are causal from $x_i=\beta(t_i)$ to $x_{i+1}=\beta(t_{i+1})$.     We observed that $\hat{d}_\tau$ does not always define a metric space for arbitrary $\tau$.  For example in Minkowski space if we take $\tau=t^3$, the null distance $\hat{d}_\tau(p,q)=0$ for all $p,q \in t^{-1}(0)$. 

Carlos Vega and I proved $\hat{d}_\tau$ defines a metric space when $\tau$ is a regular cosmological time in the sense of Anderson-Howard-Galloway \cite{AGH}.  
The cosmological time $\tau$ at $p$ is defined to be the supremum of the Lorentzian distance from
$q$ to $p$ over all $q$ in the past of $p$ CITE.  It is ``regular'' if it is finite on all of $M$ and converges to $0$ along all past inextensible curves \cite{AGH}.  One may envision examples like big bang spacetimes and maximal future developments from some initial data sets as possible examples of spaces with regular cosmological time functions.    The great advantage of using $\hat{d}_\tau$ is that it captures causality:
\be
\textrm{$p$ is in the future of $q$ $\qquad\implies\qquad$ $\hat{d}_\tau(p,q)=\tau(p)-\tau(q)$}.
\ee
We say that $\hat{d}_\tau$ {\em encodes causality} when this is an $\iff$.   We prove that when 
$\hat{d}_\tau$ encodes causality then $\hat{d}_\tau$ is also definite and thus defines a metric space.  
We observed that $\hat{d}_\tau$ encodes causality in warped product spacetimes of the form $-dt^2 + f(t)^2 g_0$ where $\tau=t$.  Indeed the balls in such spaces are shaped like cylinders around the lightcones.
\cite{Sormani-Vega-null}
\begin{itemize}
\item{\em What spaces have regular cosmological time functions?}
\item{\em When does $\hat{d}_\tau$ encode causality?}
\end{itemize}
Wald and Yip first introduced the cosmological time function as the maximal lifetime function in \cite{Wald-Yip}.   It has since been studied by Andersson-Barbot-B\'eguin-Zeghib \cite{ABBZ}, Cui-Jin \cite{Cui-Jin},
Ebrahimi \cite{Ebrahimi}, and many others but it must be explored further.

Currently Carlos Vega and I are currently exploring the $\mathcal{SF}$ convergence of big bang spacetimes.  Recall the classic FLRW big bang spacetimes have metrics of the form $dt^2 +f^2(t) g_0$ with $t>0$ and $\lim_{t\to 0} f(t) =0$.  In such
spaces the cosmological time, $\tau=t$, and so it is regular and in fact smooth and we can define a 
metric space $(X,\hat{d}_\tau)$, which encodes causality.  We have proven that there is a single big bang point, $p_{BB}$, in the metric completion of this space, $\bar{X}$, and that cosmological time, $\tau(p)=\hat{d}_\tau(p_{BB},p)$.  We can then generalize the notion of big bang spacetime to any spacetime for which
the cosmological time function defines a metric space $(X,\hat{d}_\tau)$ which encodes causality that
has a big bang point, $B \in \bar{X}$ such that $\tau(p)=\hat{d}_\tau(B,p)$.  
\begin{itemize}
\item{\em Which Lorentzian manifolds are generalized big bang spaces?}
\end{itemize}
The pointed $\mathcal{F}$
convergence of such spaces based at the big bang points is then well defined and one has compactness theorems with limit spaces which are integral current spaces with causal structures defined using
\be
\textrm{$q_1$ is in the future of $q_2$ $\qquad\iff\qquad$ $\hat{d}_\tau(q_1,q_2)
=\hat{d}_\tau(B,q_1)-\hat{d}_\tau(B,q_2)$}.
\ee

Currently Anna Sakovich and I are exploring $\mathcal{SF}$ convergence for future maximal developments of initial data sets where the cosmological time function is regular and $\tau^{-1}(0)$ is the initial Cauchy surface.  We are examining particular sequences of black hole spacetimes whose mass is converging to $0$ to test that their metric spaces defined using the null distance do indeed converge in the intrinsic flat sense and we are formulating appropriate definitions and conjectures.  It is possible we might be able to prove a compactness theorem in this setting as well.   
\begin{itemize}
\item{\em When does a future maximal development have a cosmological time function that is $0$ on the initial Cauchy surface?}
\item{\em Which future developments of initial data sets have null distances that encode causality?}   
\item{\em Are there other canonical time functions that are more suited to study future maximal developments of the Einstein equations?}  
\end{itemize}

\bibliographystyle{plain}
\bibliography{2017}
\vspace{-1cm}
\end{document}